\documentclass[prx,aps,twocolumn,amsmath,amssymb]{revtex4-1}

\pdfoutput=1
\usepackage{hyperref}
\usepackage{graphicx}
\usepackage[usenames]{color}
\usepackage{bm}
\usepackage[final]{movie15}

\def\cal#1{\mathcal{#1}}
\def\eqq#1{Eq.~(\ref{#1})}
\def\eq#1{(\ref{#1})}
\def\av#1{\langle #1 \rangle}

\def\f#1{Fig.~\ref{#1}}

\def\c#1{~\cite{#1}}

\def\x{{\bm x}}
\def\y{{\bm y}}
\def\rr{{\bm R}}
\def\r{{\bm r}}

\def\s#1{Section~\ref{#1}}

\def\beq{\begin{equation}}
\def\eeq{\end{equation}}
\def\bea{\begin{eqnarray}}
\def\eea{\end{eqnarray}}

\begin{document}

\title{Phase separation and large deviations of lattice active matter}
\author{Stephen Whitelam$^1$}\email{{\tt swhitelam@lbl.gov}}
\author{Katherine Klymko$^2$}
\author{Dibyendu Mandal$^3$}
\affiliation{
$^1$Molecular Foundry, Lawrence Berkeley National Laboratory, 1 Cyclotron Road, Berkeley, CA 94720, USA\\
$^2$Department of Chemistry, University of California at Berkeley, Berkeley, CA 94720, USA \\
$^2$Department of Physics, University of California at Berkeley, Berkeley, CA 94720, USA}
\begin{abstract}
Off-lattice active Brownian particles form clusters and undergo phase separation even in the absence of attractions or velocity-alignment mechanisms. Arguments that explain this phenomenon appeal only to the ability of particles to move persistently in a direction that fluctuates, but existing lattice models of hard particles that account for this behavior do not exhibit phase separation. Here we present a lattice model of active matter that exhibits motility-induced phase separation in the absence of velocity alignment. Using direct and rare-event sampling of dynamical trajectories we show that clustering and phase separation are accompanied by pronounced fluctuations of static and dynamic order parameters. This model provides a complement to off-lattice models for the study of motility-induced phase separation.
\end{abstract}

\maketitle

{\em Introduction --} Active matter refers to systems whose elements propel themselves by dissipating energy. Natural examples of active matter include bacteria; synthetic examples include suspensions of colloids that can catalyze chemical reactions on their surface\c{tailleur2008statistical,weitz2015self,baskaran2009statistical,peruani2012collective,barre2015motility,peruani2011traffic,ramaswamy2010mechanics, Romanczuk_2012,redner2013structure, marchetti2013hydrodynamics, buttinoni2013dynamical, stenhammar2013continuum, Yeomans_2014, cates2015motility, Menzel_2015, Bechinger_2016}. Continuous dissipation of energy ensures that active matter is `far' from equilibrium, and able to display complex behavior that includes the generation and rectification of large fluctuations\c{weitz2015self,Narayan_2007,Wan_2008, Angelani_2009, DiLeonardo_2010, Angelani_2010, Ghosh_2013, Kaiser_2014, Mallory_2014_PRE_I}; anomalous interfacial properties\c{Bialke_2015}; and phase separation in the absence of interparticle attractions\c{tailleur2008statistical,peruani2011traffic,thompson2011lattice,PhysRevE.89.012718,Fily_2012, Palacci_2013,buttinoni2013dynamical, redner2013structure, stenhammar2013continuum, Mognetti_2013, Speck_2014, Stenhammer_2015, cates2015motility, Redner_2016}. This latter phenomenon, known as motility-induced phase separation (MIPS), is similar in some respects to equilibrium gas-liquid phase separation, e.g. MIPS can be described by free-energy-like objects\c{barre2015motility,Cates_2015,Tailleur_2008,redner2013structure}, and different in others, e.g. active clusters fluctuate more than passive ones\c{peruani2010cluster,Fily_2012,peruani2012collective} (and more generally, it may be difficult to define the concept of a nonequilibrium `phase'\c{dickman2016phase}). 

To probe these connections at a fundamental level it is natural to identify the simplest models that exhibit such phenomena. Lattice models enable us to identify the microscopic origin of emergent phenomena, and they can be simulated on larger scales than their off-lattice counterparts, so facilitating calculation of e.g. critical exponents\c{baxter1982exactly,binney1992theory, chandler1987introduction,binder1986introduction}. The Ising model is the simplest model that displays equilibrium phase separation\c{chandler1987introduction}. The Katz-Lebowitz-Spohn driven lattice gas is the prototypical example of drive-induced phase separation\c{katz1984nonequilibrium, zia2010twenty}. In active matter there exist lattice models of MIPS induced by velocity alignment\c{solon2015flocking,solon2013revisiting,peruani2011traffic,PhysRevE.89.012718}, but lattice models that account only for volume exclusion and persistent motion do not show phase separation\c{sepulveda2017wetting,soto2014run}. 

Here we introduce a lattice model that exhibits MIPS, in the absence of velocity alignment, and so allows study of the phenomenon in the simplest possible setting. Our starting point is the observation that simple kinetic arguments used to describe MIPS in off-lattice models appeal only to the fact that active particles diffuse and move persistently in a direction that fluctuates\c{redner2013structure}. We show that a lattice model of active matter that captures the essence of such motion indeed exhibits MIPS (see \f{fig1}), but only if particles possess the ability to move in a direction other than that of their drift. MIPS occurs in a region of phase space analogous to where it occurs off lattice (\f{fig2}). We also use a simple rare-event sampling method\c{klymko2017rare,lattice2} to show that clustering and phase separation is accompanied by non-Gaussian fluctuations of a particular dynamic order parameter (\f{fig3}). This model allows the study of MIPS in a simple setting, and provides a complment to other models of the phenomenon\c{tailleur2008statistical,peruani2011traffic,thompson2011lattice,PhysRevE.89.012718,Fily_2012, Palacci_2013,buttinoni2013dynamical, redner2013structure, stenhammar2013continuum, Mognetti_2013, Speck_2014, Stenhammer_2015, cates2015motility, Redner_2016}.
\begin{figure*}[t] 
    \centering
    \includegraphics[width= 0.9 \linewidth]{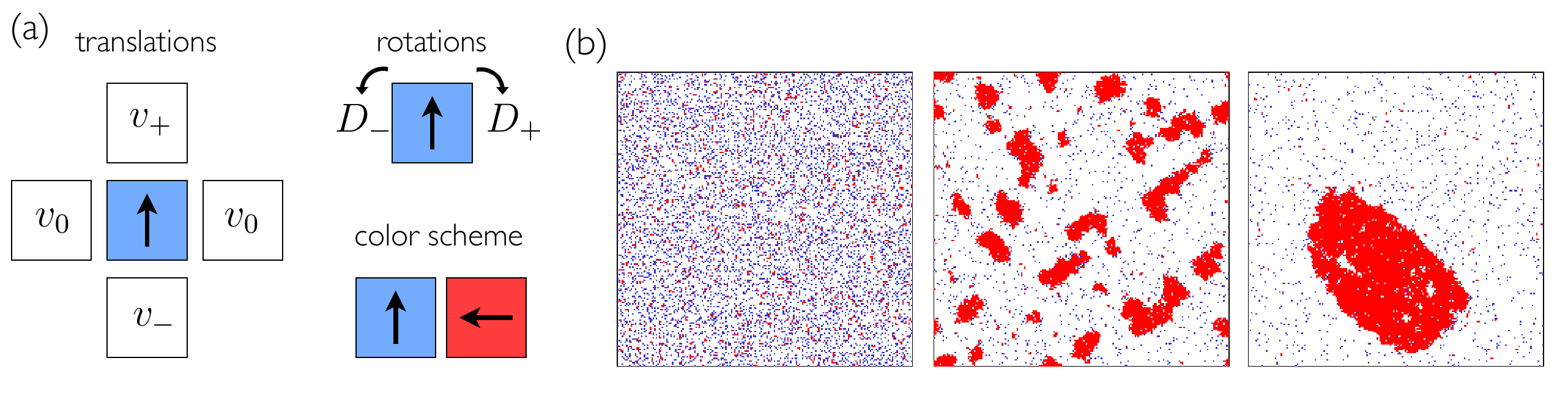} 
    \caption{(a) Rates for the motion of isolated lattice-based active particles (particles may not move to an occupied site), and the color scheme used in pictures: particles that point toward nearest-neighbor particles are shown red, and those that do not are shown blue. (b) Time-ordered configurations for density $\phi=1/5$ and $v_+ = 25$, showing motility-induced phase separation. Lattice size is $200^2$.}
    \label{fig1}
 \end{figure*} 

{\em Model --} We consider a square lattice of size $L^2$ in two dimensions, on which live $N$ hard particles. The particle density is $\phi=N/L^2$. We apply periodic boundaries in both directions.  As shown in \f{fig1}(a), particles $\alpha = 1,2,\dots,N$ possess a (unit) orientation vector ${\bm e}_\alpha$ that can point in the direction of any nearest-neighbor site. Particle $\alpha$ on site $i$ moves to a vacant nearest-neighbor site $j$ with rate $v_+$, $v_-$, or $v_0$, if ${\bm e}_\alpha \cdot {\bm r}_{ij} =+1,-1$, or 0, respectively, where ${\bm r}_{ij}$ is the unit vector pointing from site $i$ to site $j$. Particles cannot move to an occupied site. A particle's orientation vector rotates $\pi/2$ clockwise with rate $D_+$, and $\pi/2$ counter-clockwise with rate $D_-$. The orientation of a particle is unaffected by the orientation of neighboring particles (c.f. Refs.\c{solon2013revisiting,peruani2011traffic,PhysRevE.89.012718}). We simulated collections of particles using a continuous-time Monte Carlo algorithm\c{gillespie2005general}. We choose any possible process with probability $W/R$, where $W$ is the rate of the process and $R$ the sum of rates of all possible processes, and update time by an amount $1/R$ after each move. An isolated active lattice particle moves in a manner similar to that of its off-lattice active Brownian counterpart (see SI Secs. 1\&2) -- both move ballistically on short scales and diffusively on large scales --  and so we expect collections of such on-lattice particles to exhibit MIPS.

In \f{fig1}(b) we show that this expectation is borne out. Randomly dispersed and oriented particles readily cluster and undergo phase separation, here via a spinodal decomposition-like mechanism involving the aggregation of many clusters (elsewhere in parameter space we observe nucleation and growth of clusters). Similar in qualitative terms to their off-lattice counterparts~\cite{Fily_2012, redner2013structure}, clusters show pronounced fluctuations and transient internal voids: see Fig. S3. In this paper we model unbiased rotational diffusion of the orientation vector ($D_+=D_-\equiv D_{\rm rot}=1/10$). In order to mimic positional diffusion that would occur off-lattice, we allow lateral and backward motion with some rate that is in general less than the drift rate (we set $v_-=v_0 = 1$). The presence of such motion is crucial. When $v_0 = v_-=0$, i.e. when particles can move only in the direction of alignment, phase separation does not occur\c{sepulveda2017wetting,soto2014run} (see Fig. S4). Two particles that meet head-on cannot move until one of them rotates. When jammed in this way they cannot merge with larger clusters in order to drive phase separation. This effect does not occur in off-lattice models or experiment, where two agents that meet head-on can slip past each other (by rectifying each other's motion). In other words, lattice-based active particles that cannot move against their orientation vector experience an unphysical kinetic trap that prevents MIPS (an exception is the model of Ref.\c{thompson2011lattice}, which achieves phase separation on-lattice by using a coarse-grained density field, effectively allowing particles to pass through each other). Introduction of local diffusion (nonzero $v_-,v_0$) removes this trap. Thus, in the absence of velocity alignment, MIPS on-lattice is achieved by a combination of volume exclusion, persistent motion, and local diffusive motion.

In \f{fig2} we show in a space of density $\phi$ and the rate $v_+$ for forward motion where MIPS occurs. As as a simple measure of clustering we use $f_4$, the fraction of particles with 4 neighbors. \f{fig2} shows the mean $\av{f_4}$ and log-variance $\ln(\av{f_4^2}-\av{f_4}^2)$ of this quantity from single simulations begun from disordered initial conditions (we defined averages of a microstate-dependent quantity $Q(C)$ as $\av{Q} =\sum_k Q(C_k)R(C_k)^{-1}/\sum_k R(C_k)^{-1}$, where $k$ labels microstates and $1/R(C)$ is the mean time taken to escape microstate $C$). The region in which phase separation occurs can be predicted by a flux-balance argument (see SI Sec. 3) similar to that used off-lattice\c{redner2013structure}. From this we estimate that a fraction
\beq
\label{fb_main}
 f=\frac{{\rm Pe} \, \kappa-1/\phi}{{\rm Pe} \, \kappa-1}
\eeq
of particles will be in the dense phase, where $4 \kappa \equiv 1-(1-2D_{\rm rot}/\Sigma)^{2 \Sigma/D_{\rm rot}}$; $\Sigma \equiv v_++v_-+2v_0+2D_{\rm rot}$; and the P\'eclet number ${\rm Pe} \equiv (v_+-v_-)/(2D_{\rm rot})$ (for the parameters used we have ${\rm Pe} = 5(v_+-1)$). In \f{fig2} we plot the line $f=1/2$. This line matches approximately the curvature of the phase boundary obtained by computer simulation, confirming that MIPS on-lattice occurs for a density-dependent P\'eclet number, as it does off lattice\c{redner2013structure}. In addition, the variance of $f_4$ is large even in the ordered phase, indicating pronounced fluctuations of clusters (which is the case off lattice\c{Fily_2012}).

\begin{figure*}[] 
    \centering
\includegraphics[width=0.9\linewidth]{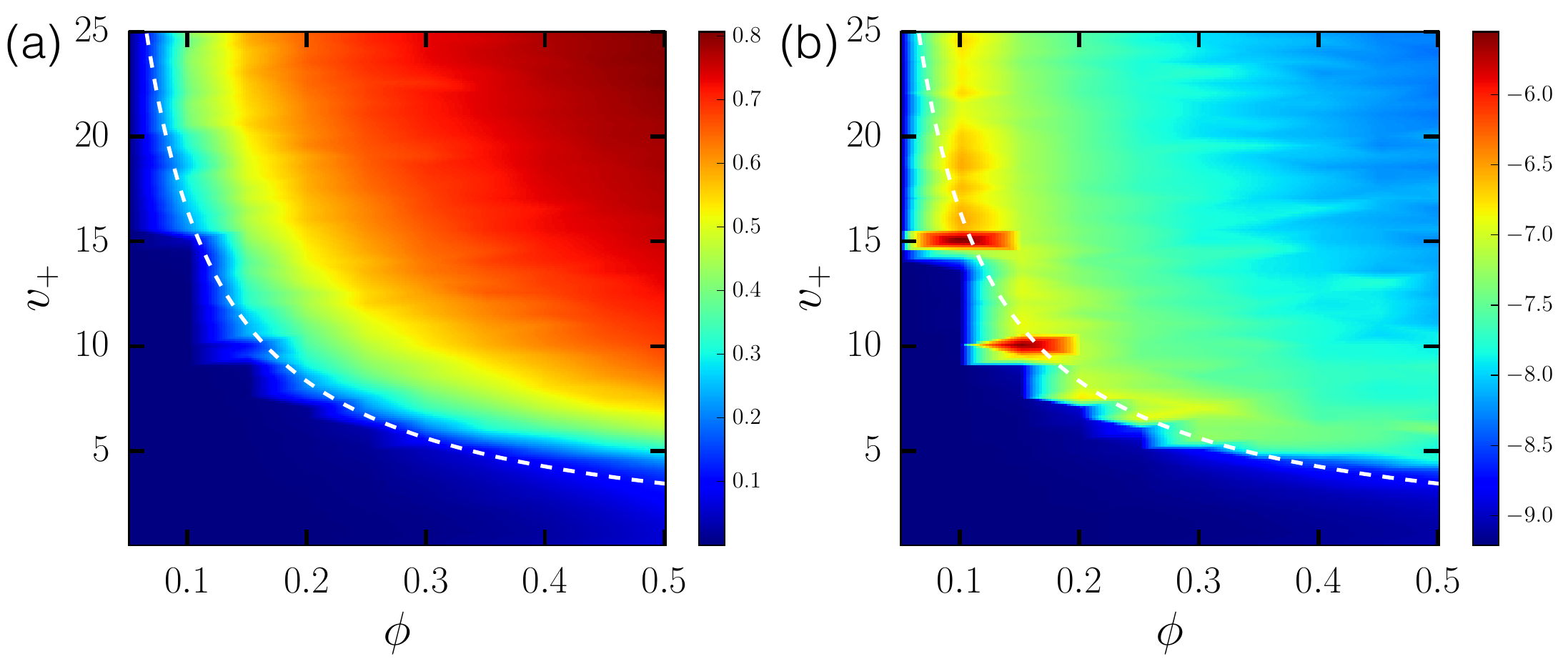} 
    \caption{(a) Mean and (b) logarithm of the variance of $f_4$, the fraction of particles with 4 neighbors, as a function of density $\phi$ and $v_+$ (the P\'{e}clet number ${\rm Pe} = 5(v_+-1)$). As in the off-lattice model of Ref.\c{redner2013structure}, we observe phase separation at a density-dependent value of Pe. The dotted white line is the contour $f=1/2$ from \eqq{fb_main}. Lattice size is $100^2$.}
    \label{fig2}
 \end{figure*} 

{\em Trajectory sampling --} To more thoroughly probe the fluctuations associated with clustering and phase separation we used a simple method of rare-event sampling\c{klymko2017rare,lattice2} motivated by the `thermodynamics of trajectories' or `$s$-ensemble' formalism\c{ruelle2004thermodynamic, garrahan2007dynamical, Lecomte2007, garrahan2009first, hedges2009dynamic}. Briefly, we wish to calculate $\rho(a,K)$, the probability distribution, over an ensemble of trajectories, of a quantity $a=A/K$, where $A$ is an observable extensive in the length of the trajectory and $K$ is the number of simulation steps (configuration changes) in each trajectory in the ensemble. We define a trajectory as a sequence $\x = \{C_1,C_2,\dots,C_K\}$ of microstates $C_k$ visited by the dynamics. The probability of a step $C_k \to C_{k+1}$ in this sequence is $W(C_k \to C_{k+1})/R(C_k)$, where $W(C_k \to C_{k+1})$ is the rate for the enacted process, and $R(C_k) \equiv \sum_{C'} W(C_k \to C')$ is the sum of rates of all processes leading out of state $C_k$. Direct simulation of the model allows for efficient sampling of $\rho$ for typical values of $a$, and poor sampling of $\rho$ for rare values of $a$. We therefore make use of a `change of measure'\c{touchette2009large, maes2008canonical, giardina2011simulating, lecomte2007numerical, nemoto2014computation, jack2015effective, nemoto2016population}, and introduce a reference model whose rates
\beq
\label{rm}
W_{\rm ref}(C \to C') = {\rm  e}^{-s \alpha(C \to C')} W(C \to C')
\eeq
are chosen so that the reference model's typical values of $a$ are generated with low probability by the original model. Here $\alpha(C \to C')$ is the change of $A$ upon moving from $C$ to $C'$, and $s$ is a parameter. The likelihood $w[\x]$ that a trajectory $\x$ generated by the reference model {\em would} have been generated by the original model is 
\beq
w[\x] =e^{s A[\x]} \prod_{k=0}^{K-1} \frac{R_{\rm ref}(C_k)}{R(C_k)},
\eeq
with $R_{\rm ref}(C_k) \equiv \sum_{C'} W_{\rm ref}(C_k \to C')$. The quantity we want, the probability density of $a$ over trajectories of length $K$, is given by
\bea
\label{com}
\rho(a,K)= \sum_\x{P_{\rm ref}[\x] w[\x] \delta{(A[\x]-K a)}},
\eea
a sum over values of $w$ for trajectories of the reference-model dynamics, each of which possesses a given value of $A$ (namely, $ A = a K$). Here $P_{\rm ref}[\x]$ denotes the probability with which trajectory $\x$ is generated by the reference model, and $w[\x]=P[\x]/P_{\rm ref}[\x]$ is the ratio of that quantity and the corresponding quantity for the original model. The quantity $I(a) \equiv  -K^{-1} \ln \rho(a,K)$ is the large-deviation rate function for $a$ (in the limit of large $K$), which quantifies the likelihood of departures, small and large, from typical behavior\c{touchette2009large}. 

We can obtain a bound $I_0(a)>I(a)$ on the rate function from individual trajectories of the reference model\c{klymko2017rare,lattice2}; this bound is
\beq
\label{rf}
I_0(a_s) =-s a_s - \sum_C \pi_{\rm ref}(C) \ln \frac{R_{\rm ref}(C)}{R(C)},
\eeq
where $a_s = A_s/K$ is a value of $a$ typical of the reference model, and $\pi_{\rm ref}(C)$ is the steady-state probability of visitation, by the reference model, of state $C$.
For a given value of $s$ we possess a reference model with typical value $a_s$ for the observable $A/K$. This model, through \eqq{rf}, yields one point $I_0(a_s)$ on the curve $I(a)$; repeating the procedure for several values of $s$ gives the whole curve.

\begin{figure*}[t] 
    \centering
    \includegraphics[width=0.9 \linewidth]{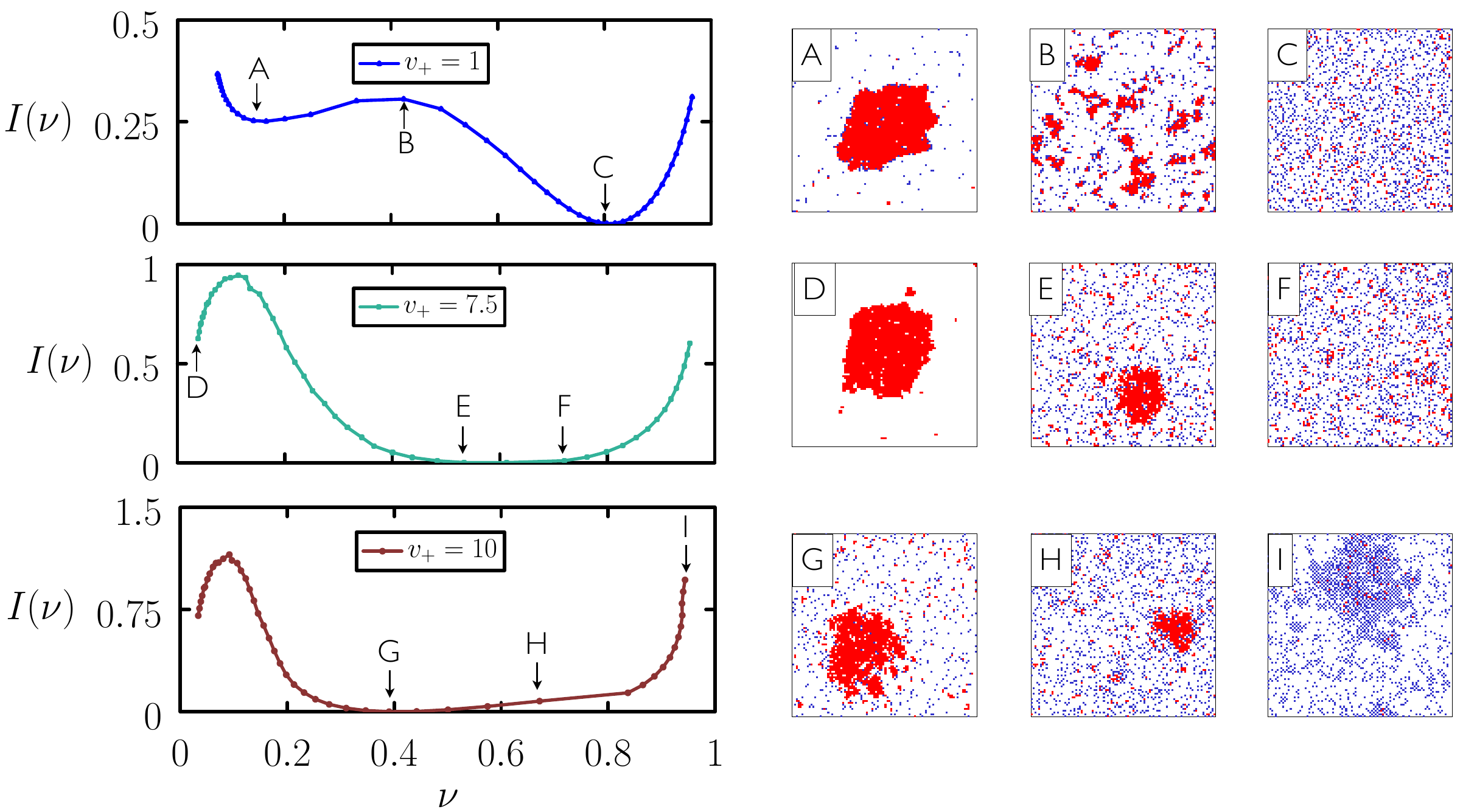} 
    \caption{Trajectory sampling yields an upper bound $I_0(\nu)$ on the large-deviation rate function for trajectory activity $\nu$. As we vary $v_+$ we change typical trajectories ($I=0)$ from being more active (large $\nu$, top) to being less active (small $\nu$, bottom), via an intermediate region (middle). Configurations from some typical and rare trajectories are shown right, indicating that activity is strongly correlated with clustering. Lattice size is $100^2$, and density is $\phi=1/5$.}
    \label{fig3}
 \end{figure*} 

We choose an observable $a$ guided by studies of glasses. There, authors often choose to count the number of events that occur in a particular time, with the average time taken to leave configuration $C$ being $1/R(C)$. This choice allows the identification of phase transitions out of equilibrium\c{garrahan2009first,hedges2009dynamic,budini2014fluctuating}. Similar physics should be accessible by measuring the values of $R(C)$ of states explored by a fixed number of configuration changes. We therefore take the reference-model bias $\alpha(C\to C')=B(C')$, where $B(C') = R(C')$ is the escape rate from the state to which the model is moving (we use $B$ to emphasize that this choice can be varied); the order parameter against which dynamics is conditioned is $ A / K = K^{-1} \sum_{k=0}^{K-1} B(C_{k+1})$, the mean relaxation rate of configurations comprising the trajectory (hereafter called the `activity' of the trajectory\c{garrahan2009first}). The reference model \eq{rm} is then $W_{\rm ref}(C \to C') = {\rm  e}^{-s B(C')} W(C \to C')$. We simulate it as we do the original model, but now choosing events $C \to C'$ with probabilities $W_{\rm ref}(C \to C')/R_{\rm ref}(C)$. One can guide the reference model toward quickly- or slowly-relaxing configurations depending upon the sign and magnitude of $s$. We compute \eq{rf} by running a single reference-model trajectory, for a given value of $s$, and evaluating the expression
\begin{eqnarray}
\label{rf2}
I_0(a_s) & = & -s K^{-1}\sum_{k=0}^{K-1} [B(C_{k+1})-B(C_k)]  \\
& & - K^{-1}\sum_{k=0}^{K-1} \ln \frac{\sum_{C'}{\rm e}^{-s [B(C')-B(C_k)]} W(C_k \to C')}{\sum_{C'}W(C_k \to C')}; \nonumber
\end{eqnarray}
note that we have written $\ln \sum_{C'}{\rm e}^{-s B(C')} \equiv s B(C)+\ln \sum_{C'}{\rm e}^{-s [B(C')-B(C)]}$, so that numbers appearing in exponentials are not too large. The first term on the right-hand side of \eq{rf2} becomes negligible for large $K$. 

 The reference model is a tool whose purpose is to tell us with what probability the original model will yield (rare) values of an observable. At the same time, configurations of the reference model indicate the nature of the configurations that will be visited, with low probability, by the original model. The reference model satisfies the relation $W_{\rm ref}(C\to C')/W_{\rm ref}(C'\to C)={\rm e}^{-s[B(C')-B(C)]} W(C\to C')/W(C'\to C)$. Given that $B=R$ counts the numbers and types of particle-vacancy contacts, it is clear that biasing the system toward quickly- ($s>0$) or slowly-relaxing ($s<0$) configurations is akin to equipping particles with (anisotropic) repulsions or attractions, respectively. In the case $v_+=v_-=v_0$ the original model comprises a set of diffusive hard particles, and the reference model, which measures the number of particle-vacancy bonds, is the Ising lattice gas with particle-particle interaction energy $-s$. Rare, slowly-relaxing configurations of the original model therefore look like typical lattice gas configurations in the presence of an attractive interaction -- i.e. they can be phase-separated -- and rare, quickly-relaxing configurations of the original model look like typical configurations of the lattice gas in the presence of repulsive interactions (which for certain particle densities are periodic and so hyperuniform\c{jack2015hyperuniformity}). 
 
In \f{fig3} we show $I_0(\nu)$, an upper bound on the rate function associated with the (scaled) activity $\nu \equiv (\Sigma N K)^{-1} \sum_{k=0}^{K-1} B(C_k)$ (here $N$ is the number of particles). To make this figure we ran single reference-model trajectories of length $K = {\cal O}(10^8)$ for several values of $s$, both positive and negative. For each trajectory (prepared by starting with a single cluster and running until we entered steady state) we evaluated $a$ ($=\nu N \Sigma$) and $I_0(a)$, using \eqq{rf2}. Large values of $\nu$ correspond to active trajectories, i.e. those that whose configurations change rapidly. Typical behavior is signaled by $I=0$\c{touchette2009large}, while atypical behavior corresponds to large values of $I$.

We see that increasing $v_+$ changes the typical behavior of the system from being more active and disordered (top) to being less active and clustered (bottom), via an intermediate regime (middle)~\footnote{The linguistically unfortunate result being that making passive particles ($v_+=1$) active ($v_+>1$) reduces their emergent activity ($\nu$ decreases).}, consistent with the transition from disordered to phase-separated configurations shown in \f{fig2}. In addition, the bimodality seen in the rate functions indicates the presence of a transition in terms of $s$, the field conjugate to $\nu$\c{lattice2}. Thus we see transitions at the level of typical and atypical trajectories, a scenario similar to that seen in model lattice proteins\c{mey2014rare}. The snapshots shown in the figure indicate that activity ($\nu$) and clustering are strongly (but not perfectly) correlated: in general, less-active trajectories contain clustered configurations and active trajectories contain non-clustered configurations, but there also exist rare, active trajectories that exhibit `checkerboard' clustering. In the lower two panels the minima of $I$ are broad, indicating the existence of pronounced fluctuations. While we expect fluctuations near a phase boundary\c{binney1992theory}, it is notable that strongly non-Gaussian fluctuations persist into the region of phase separation (bottom panel). Non-Gaussian fluctuations of cluster size are seen in off-lattice models of active matter\c{Fily_2012}.
 
{\em Conclusions --} We have presented an on-lattice model of hard active particles that exhibits MIPS in the absence of velocity alignment. The model exhibits clustering and phase separation qualitatively similar to that seen in off-lattice models. Both direct simulations and trajectory-sampling methods show that pronounced fluctuations are present even within the ordered phase of the system. Lattice models provide a simple complement to off-lattice models, and the one presented here provides a simple way of studying motility-induced phase separation in the absence of velocity alignment. 

{\em Acknowledgments -- } We thank Michael Hagan and Juan Garrahan for discussions. SW performed work at the Molecular Foundry, Lawrence Berkeley National Laboratory, supported by the Office of Science, Office of Basic Energy Sciences, of the U.S. Department of Energy under Contract No. DE-AC02--05CH11231. KK acknowledges support from an NSF Graduate Research Fellowship. DM acknowledges financial support by the U. S. Army Research Laboratory and the U. S. Army Research Office under contract W911NF-13-1-0390.

%\bibliography{bib_v2}

%merlin.mbs apsrev4-1.bst 2010-07-25 4.21a (PWD, AO, DPC) hacked
%Control: key (0)
%Control: author (8) initials jnrlst
%Control: editor formatted (1) identically to author
%Control: production of article title (-1) disabled
%Control: page (0) single
%Control: year (1) truncated
%Control: production of eprint (0) enabled
%

%\onecolumngrid
\clearpage

\renewcommand{\theequation}{S\arabic{equation}}
\renewcommand{\thefigure}{S\arabic{figure}}
\renewcommand{\thesection}{S\arabic{section}}

\setcounter{equation}{0}
\setcounter{section}{0}
\setcounter{figure}{0}

\setlength{\parskip}{0.25cm}%
\setlength{\parindent}{0pt}%

%\appendix  
 
%\title{Supplemental information for ``Phase separation and large deviations of lattice active matter''}
%\author{Stephen Whitelam$^1$}\email{{\tt swhitelam@lbl.gov}}
%\author{Katherine Klymko$^2$}
%\author{Dibyendu Mandal$^3$}
%\affiliation{
%$^1$Molecular Foundry, Lawrence Berkeley National Laboratory, 1 Cyclotron Road, Berkeley, CA 94720, USA\\
%$^2$Department of Chemistry, University of California at Berkeley, Berkeley, CA 94720, USA \\
%$^2$Department of Physics, University of California at Berkeley, Berkeley, CA 94720, USA}

%\maketitle

\section{Motion of an isolated off-lattice active Brownian particle}
\label{offlat}
 
In this section we recall some features of the typical motion of an isolated off-lattice two-dimensional active Brownian particle, of the type considered in some simulation studies\c{Fily_2012,redner2013structure}. In \s{onlat} we show that an on-lattice active particle moves in a qualitatively similar way. The situation and notation considered in this section draws upon Section 4.3.1 of Ref\c{romanczuk2012active} (although is not identical to the situation considered there); more comprehensive treatments of active-particle motion can be found elsewhere\c{ten2011brownian,romanczuk2012active,howse2007self}. 
 
\subsection{Preliminaries}
Consider numerical integration of the position of an active Brownian particle in $d=2$. The particle is subject to thermal fluctuations and able to move deterministically in the direction of its orientation vector. After step $N$ of the simulation its position vector is $\rr_N = \sum_{i=1}^N \Delta \r_i$, where
\bea
\label{r1}
\Delta \r_i &=& \ell_0 \left( \cos \theta_{i-1} \hat{\x}+\sin \theta_{i-1} \hat{{\bm y}}\right)\nonumber \\
&+& \sqrt{2 \Delta \tau D_x}\eta^x_i \hat{\x}+\sqrt{2 \Delta \tau D_y}\eta^y_i \hat{{\bm y}}.
\eea
Here $\hat{{\bm x}}$ and $\hat{{\bm y}}$ are Cartesian unit vectors; $\ell_0 = V_0 \Delta \tau$ is the displacement magnitude of the deterministic force; $\Delta \tau$ is the integration timestep; $D_x$ and $D_y$ are diffusion constants; $\theta_{i-1}$ is the angle (after step $i-1$ and before step $i$) between the particle's orientation vector and the $x$-axis; and $\eta^{x}_i$ and $\eta^{y}_i$ are Gaussian white noise terms with zero mean and unit variance, i.e. $\av{\eta_i^\alpha}=0$ and $\av{\eta^\alpha_i\eta^\beta_j} = \delta_{\alpha,\beta} \delta_{i,j}$.

Let the particle angle evolve according to
\beq
\label{r1b}
\theta_i = \theta_{i-1}+\eta_i,
\eeq
where $\eta_i$ is drawn from an even distribution $P(\eta_i)$. Let this distribution be bounded by $\pm \pi$ and have zero mean, in which case $\av{ \sin \eta_i} = 0$ and $\lambda \equiv \av{\cos \eta_i} \neq 0$ (in general). We assume that angular changes at different times are uncorrelated.

For the sake of generality we shall consider three cases. The first is that of {\em driven matter} (see e.g. Refs.\c{katz1984nonequilibrium, glanz2012nature}), where the particle's orientation vector does not rotate. In this case $P(\eta_i) = \delta(\eta_i)$ and $\lambda = 1$. The second case is that of {\em active matter}, in which the particle's orientation angle rotates diffusively. In this case we have $0 < \lambda <1$. For the particular case of a Gaussian distribution $P(\eta_i)$ we have
\bea
\lambda &=& \int_{-\pi}^{\pi} {\rm d} \eta \cos \eta \cdot \frac{1}{\sqrt{2 \pi \sigma^2}} \exp\left( -\frac{\eta^2}{2 \sigma^2}\right) \nonumber \\
&\approx& {\rm e}^{-\sigma^2/2},
\eea
for $\sigma$ small enough that the limits of the integral can be approximated by $\pm \infty$ (the exact solution can be written in terms of the error function). The third case is that of {\em Brownian matter}, where $P(\eta_i)$ is drawn uniformly from the interval $[-\pi, \pi)$. In this case $\lambda = 0$ (here the particle moves diffusively, with a diffusion constant renormalized by the drift parameter: see e.g. Ref.\c{tanaka2017hot}). 

To work out properties of the particle's motion we will need 
\bea
\label{rec1}
\av{\cos \theta_i} &=& \av{\cos( \theta_{i-1}+\eta_i)} \nonumber \\
&=& \av{\cos \theta_{i-1}} \av{\cos \eta_i } -\av{\sin  \theta_{i-1} } \av{\sin \eta_i }\nonumber \\
&=& \av{\cos  \theta_{i-1}} \lambda, 
\eea
using the fact that noise terms at different times are uncorrelated. \eqq{rec1} is a recursion relation and implies
\beq
\label{cos_eq}
\av{\cos \theta_i} = \cos {\theta_0} \lambda^i,
\eeq
where $\theta_0$ is the particle's initial angle. Similarly, $\av{\sin \theta_i} = \sin {\theta_0} \lambda^i$.

To compute second moments of position we need to average
\begin{widetext}
\bea
\Delta \r_i \cdot \Delta \r_j &=& \left( \ell_0 \cos \theta_{i-1} + \sqrt{2 \Delta \tau D_x} \eta^x_i\right)\left( \ell_0 \cos \theta_{j-1} + \sqrt{2 \Delta \tau D_x} \eta^x_j\right) \nonumber \\
&+& \left( \ell_0 \sin \theta_{i-1} + \sqrt{2 \Delta \tau D_y} \eta^y_i\right)\left( \ell_0 \sin \theta_{j-1} + \sqrt{2 \Delta \tau D_y} \eta^y_j\right).
\eea
\end{widetext}
Anything linear in $\eta^x$ or $\eta^y$ will not survive the averaging; what remains to be averaged is
\bea
\ell_0^2 \cos(\theta_{j-1} - \theta_{i-1}) + 2 \Delta \tau D_x \eta^x_i \eta^x_j+2 \Delta \tau D_y \eta^y_i \eta^y_j.
\eea
For $i=j$ we have
\beq
\av{\Delta \r_i \cdot \Delta \r_i} = \ell_0^2 +2(\Delta \tau D_x+\Delta \tau D_y).
\eeq
For $j>i$ we have
\bea
\av{\Delta \r_i \cdot \Delta \r_j} &=& \ell_0^2 \av{\cos(\theta_{j-1} - \theta_{i-1})} \nonumber \\
&=& \ell_0^2 \lambda^{j-i}.
\eea

\subsection{Character of motion}

The position of the particle after step $N$ is $\rr_N = \sum_{i=1}^N \Delta \r_i$, and so
\bea
\label{r_av}
\av{\rr_N}  &=& \sum_{i=1}^N \av{\Delta \r_i} \nonumber \\
\label{r_av_mid}
&=& \ell_0 \left( \cos \theta_0 \hat{\x}+\sin \theta_0 \hat{{\bm y}}\right) \sum_{i=1}^N \lambda^{i-1} \\
&=&\ell_0 {\bm e}_0 \frac{1-\lambda^N}{1-\lambda},
\eea
where ${\bm e}_0 \equiv \cos \theta_0 \hat{\x}+\sin \theta_0 \hat{{\bm y}}$ is the initial orientation vector of the particle. Here the angle brackets denote an average over trajectories (i.e. noise), for particles that start at the origin with angle $\theta_0$. 

For $\lambda \lesssim 1$ and $N$ small we can write $\lambda^N \approx 1+N \ln \lambda$, in which case 
\beq
\av{\rr_N} \approx N \ell_0 {\bm e}_0 \frac{(-\ln \lambda)}{1-\lambda},
\eeq
i.e. on small scales the particle moves ballistically. For large $N$ the mean displacement does not vanish, but tends instead to the limit
\beq
\label{r_infty}
\av{\rr_\infty}= \ell_0 {\bm e}_0 \frac{1}{1-\lambda}.
\eeq
Thus drift on short times generates a net displacement that is `remembered' by the particle at long times. Only if we average over initial orientations ${\bm e}_0$ does the net displacement vanish.  

Results for the case of driven matter $(\lambda = 1)$ and Brownian matter $(\lambda=0)$ can be obtained straightforwardly from \eq{r_av_mid}. Collecting these results we have
\bea
  \av{\rr_N} = \ell_0 {\bm e}_0 \times \left\{ 
  \begin{array}{cc} % brackets may be (...), [...], \{...\}, or left out
      N  &(\lambda = 1)\\
     \frac{1-\lambda^N}{1-\lambda} &(0< \lambda <1) \\
       0  &(\lambda = 0),
   \end{array}
   \right.
   \eea
   for off-lattice driven, active, and Brownian matter, respectively.
      
The noise-averaged mean-squared displacement is
 \begin{widetext}
 \bea
 \label{msd_active}
 \av{\rr_N \cdot \rr_N } &=& \av{\sum_{i=1}^N \Delta {\bm r }_i \cdot \sum_{i=j}^N\Delta \r_j} \nonumber \\
&=&\sum_{i=1}^N \av{\Delta \r_i \cdot \Delta \r_i} + 2 \sum_{i=1}^{N-1} \sum_{j=i+1}^N \av{\Delta \r_i \cdot \Delta \r_j} \nonumber \\
&=&N\ell_0^2 +2N(\Delta \tau D_x+\Delta \tau D_y) +2 \ell_0^2\sum_{i=1}^{N-1} \sum_{j=i+1}^N \lambda^{j-i}\nonumber \\
&=&N\ell_0^2 +2N(\Delta \tau D_x+\Delta \tau D_y) +2 \ell_0^2 \frac{\lambda}{(1-\lambda)^2} (N(1-\lambda)-1 + \lambda^N  ).
 \eea
  \end{widetext}
In the case of active matter we have ballistic motion on small scales, when $\lambda^N \approx 1$:
 \beq
  \av{\rr_N \cdot \rr_N } \approx \ell_0^2 N^2 +2N(\Delta \tau D_x+\Delta \tau D_y).
 \eeq 
We have diffusive motion on large scales (when $N \to \infty$), with an effective diffusion constant
 \bea
 D_{\rm eff} &\equiv& \lim_{N \to \infty} \frac{1}{4N \Delta \tau} \av{\rr_N \cdot \rr_N } \\  \label{deff1}
&=& \frac{\ell_0^2}{4 \Delta \tau} \frac{1+\lambda}{1-\lambda} + \frac{1}{2}( D_x+ D_y),
 \eea
 which is renormalized by the self-propulsion of the particle.
   \begin{figure*}[t!] %  figure placement: here, top, bottom, or page
    \centering
    \includegraphics[width=\linewidth]{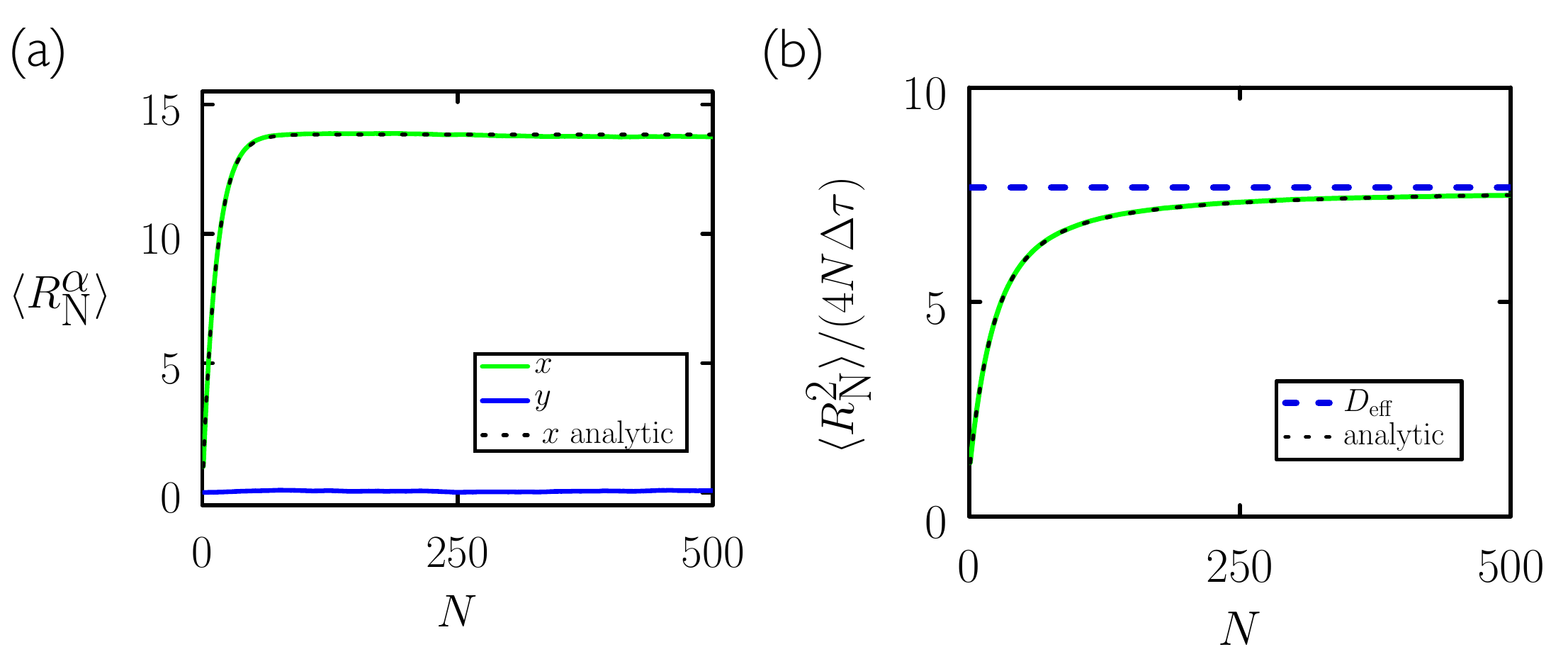} 
    \caption{Numerical integration of \eqq{r1} and \eqq{r1b} confirms the analytic results derived here, for (a) mean displacement (for $\alpha=x,y$) and (b) mean-squared displacement. The black dotted lines are analytic results \eq{r_av} and \eq{msd_active}; the blue dashed line $D_{\rm eff}$ is the result \eq{deff1}. Here $V_0 = D_x = D_y = \Delta \tau =1$, and $\sigma^2= 0.15$, which gives $\lambda \approx 0.93$. Numerical averages are taken over $10^6$ trajectories of a particle initially at the origin and oriented in the $x$-direction.}
    \label{figs1}
 \end{figure*} 
 
 Collecting results for driven, active, and Brownian matter we have
 \bea
   \av{\rr_N \cdot \rr_N }=  \left\{ 
  \begin{array}{cc} % brackets may be (...), [...], \{...\}, or left out
      N^2 \ell_0^2+2N(\Delta \tau D_x+\Delta \tau D_y)  &(\lambda = 1)\\
   {\rm  \eqq{msd_active}} &(0< \lambda <1)\\
       N \ell_0^2 + 2N (\Delta \tau D_x+\Delta \tau D_y)  &(\lambda = 0),
   \end{array} \nonumber
   \right.
   \eea
   for off-lattice driven, active, and Brownian matter, respectively.
   
 In \f{figs1} we confirm these analytic results numerically: an active Brownian particle moves ballistically at short times, possesses a mean displacement that is non-vanishing, and is effectively diffusive at long times.

 \section{Motion of an isolated on-lattice active Brownian particle}
\label{onlat}

In this section we show that the motion of an isolated lattice-based active particle is similar to that of the off-lattice particle of \s{offlat}.

\subsection{Preliminaries}

Consider an isolated on-lattice active Brownian particle of the type described in the main text, evolved using a continuous-time Monte Carlo algorithm. After step $N$ of a simulation the particle's position vector is $\rr_N = \sum_{i=1}^N \Delta \r_i$, where
\bea
\Delta \r_i &=& (p_1(i) - p_4(i)) \cos\theta_{i-1}\hat{\x}  \nonumber \\&+& (p_3(i) - p_2(i)) \sin\theta_{i-1} \hat{\x} \nonumber\\
&+&  (p_1(i) - p_4(i)) \sin\theta_{i-1} \hat{\y} \nonumber\\&+& (p_2(i) - p_3(i)) \cos\theta_{i-1} \hat{\y}. \hspace{1cm}
\eea
Here $\theta_{i-1} \in \{0, \pi/2,\pi,3 \pi/2 \}$ is the orientation angle of the particle immediately prior to step $i$. The functions $p_\alpha(i)$ have the following properties: $p_1(i)$ is 1 if $0 < \xi_i  \leq v_+/\Sigma$ and 0 otherwise, where $\xi_i$ is a random variable uniformly distributed on $(0,1]$, and $\Sigma \equiv v_++2 v_0 + v_- +2 D_{\rm rot}$ is the total rate for all the processes accessible to an isolated particle. Similarly, $p_2(i)$ is 1 if $v_+/\Sigma < \xi_i  \leq (v_++v_0)/\Sigma$ and 0 otherwise; $p_3(i)$ is 1 if $(v_++v_0)/\Sigma < \xi_i  \leq (v_++2v_0)/\Sigma$ and 0 otherwise; and $p_4(i)$ is 1 if $(v_++2v_0)/\Sigma < \xi_i  \leq (v_++2v_0+v_-)/\Sigma$ and 0 otherwise. 

The angular degree of freedom $\theta$ evolves according to $\theta_i = \theta_{i-1} + \Delta \theta_i$, where
\beq
\Delta \theta_i = p_5(i) \frac{\pi}{2} - p_6(i) \frac{\pi}{2}.
\eeq
Here $p_5(i)$ is 1 if $(v_++2v_0+v_-)/\Sigma < \xi_i \leq (v_++2v_0+v_-+D_{\rm rot})/\Sigma$ and 0 otherwise, and $p_6(i)$ is 1 if $(v_++2v_0+v_-+D_{\rm rot})/\Sigma < \xi_i \leq 1$ and 0 otherwise.

Noise terms are different times are uncorrelated and so averages $\av{\cdot}$ over noise for trajectories of $N$ total steps are given by $\prod_{i=1}^N \int_{0}^1 {\rm d} \xi_i ( \cdot )$. We then have $ \av{p_1(i) \cos \theta_{i-1}} =  \av{p_1(i)} \av{ \cos \theta_{i-1}}$ etc. We also have
\bea
\label{rr}
\av{\cos \theta_i} &=& \av{\cos \left(\theta_{i-1} + \Delta \theta_i\right)} \nonumber \\
&=& \av{\cos \theta_{i-1}} \av{\cos \Delta \theta_i} - \av{\sin \theta_{i-1}} \av{\sin \Delta \theta_i}  \nonumber \\
&=&\av{\cos \theta_{i-1}} \lambda,
\eea
where $\lambda \equiv 1-2D_{\rm rot}/\Sigma$ (note that $\lambda$ in the equations of the previous section is distinct). The recursion relation \eq{rr} implies $\av{\cos \theta_i} = \cos \theta_0 \lambda^i$, where $\theta_0$ is the initial angle of the particle. Similarly, we have $\av{\sin \theta_i} = \sin \theta_0 \lambda^i$.

We then have 
\beq
\av{ \Delta \r_i} = \frac{v_+ - v_-}{\Sigma} {\bm e}_0 \lambda^{i-1} \equiv \ell_0 {\bm e}_0 \lambda^{i-1},
\eeq
where ${\bm e}_0 \equiv (\cos \theta_0, \sin \theta_0)$ is the initial orientation vector of the particle. We have defined $\ell_0 \equiv (v_+-v_-)/\Sigma$ (note that $\ell_0$ in the equations of the previous section is distinct).  Finally,
\bea
\av{ \Delta \r_i \cdot \Delta \r_i} &=& \av{ (p_2(i)-p_3(i))^2+ (p_1(i)-p_4(i))^2} \nonumber \\&=& \frac{v_++2 v_0 + v_-}{\Sigma},
\eea
and, for $j>i$,
\bea
\av{ \Delta \r_i \cdot \Delta \r_j}&=&\av{p_1(j)-p_4(j)} \nonumber \\&\times& \av{(p_1(i)-p_4(i)) \cos(\theta_{j-1}-\theta_{i-1})}\\
&=&\ell_0^2 \lambda^{j-i-1}.
\eea

\subsection{Character of motion}

Using the results of the previous section we have
\beq
\av{\rr_N} = \sum_{i=1}^N \av{\Delta \r_i} = \frac{v_+-v_-}{\Sigma} {\bm e}_0 \frac{1-\lambda^N}{1-\lambda},
\eeq
for $0<\lambda < 1$. Recall that $\lambda \equiv 1-2D_{\rm rot} /\Sigma$. Thus
\bea
  \av{\rr_N} = \ell_0 {\bm e}_0 \times \left\{ 
  \begin{array}{cc} % brackets may be (...), [...], \{...\}, or left out
      N  &(\lambda = 1)\\
     \frac{1-\lambda^N}{1-\lambda} &(0< \lambda <1) \\
       0  &(\lambda = 0),
   \end{array}
   \right.
   \eea
   for on-lattice driven, active, and Brownian matter, respectively (driven matter corresponds to $D_{\rm rot} = 0$; Brownian matter corresponds to $\ell_0=0$, where $\ell_0 \equiv (v_+-v_-)/\Sigma$). For active matter we have ballistic motion for small $N$, 
   \beq
   \label{r_av2}
   \av{\rr_N} \approx N \ell_0 {\bm e}_0 \frac{(-\ln \lambda)}{1-\lambda},
   \eeq
  and long-time non-vanishing mean displacement,
  \beq
  \av{\rr_\infty}=\ell_0 {\bm e}_0 \frac{1}{1-\lambda} \equiv \frac{v_+-v_-}{2 D_{\rm rot}} {\bm e}_0,
  \eeq
  similar to the off-lattice result \eq{r_infty}. This result suggests defining the P\'eclet number ${\rm Pe} \equiv (v_+-v_-)/(2 D_{\rm rot})$, so that $\av{\rr_\infty} = {\rm Pe} \, {\bm e}_0$.
      \begin{figure*}[t!] %  figure placement: here, top, bottom, or page
    \centering
    \includegraphics[width=\linewidth]{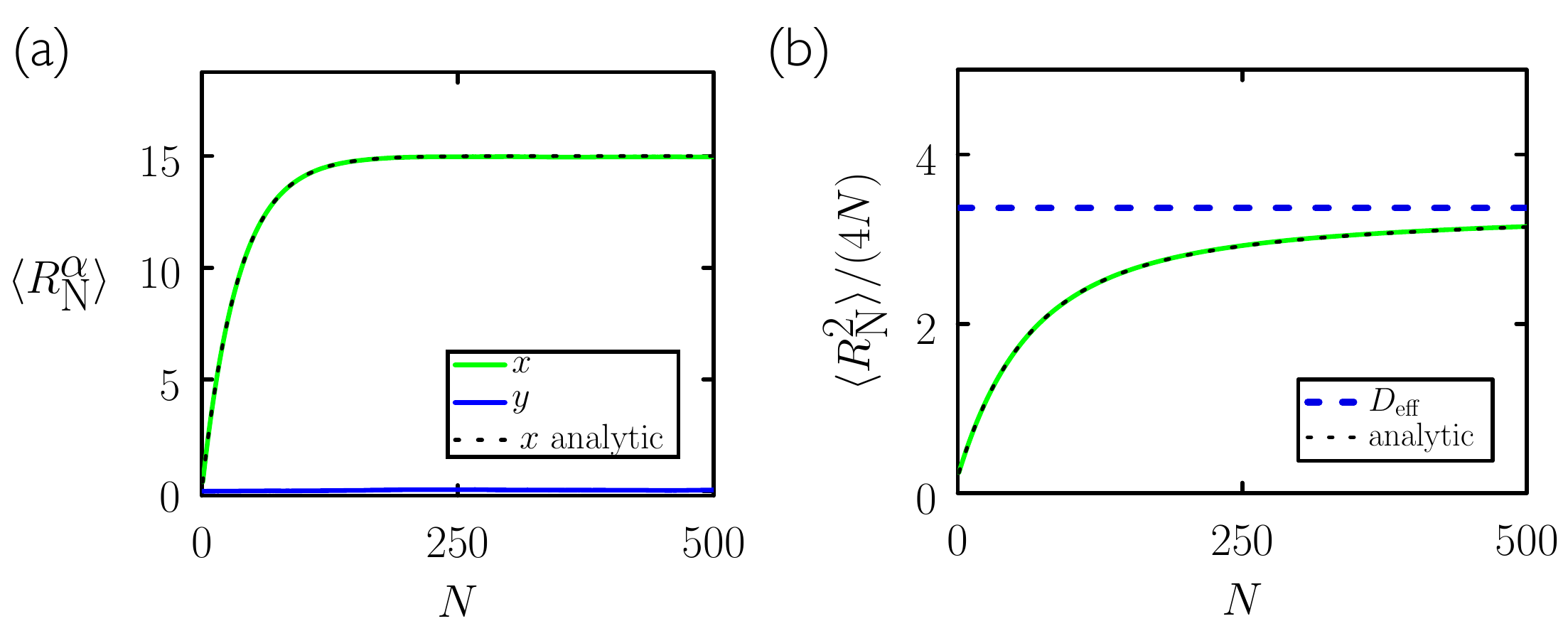} 
    \caption{Numerical evolution of an on-lattice active particle confirms the analytic results derived here, for (a) mean displacement (for $\alpha=x,y$) and (b) mean-squared displacement. The black dotted lines are analytic results \eq{r_av2} and \eq{msd_active3}; the blue dashed line $D_{\rm eff}$ is the result \eq{deff2}. Here $v_+=4$, $v_-=v_0=1$, and $D_+=D_-=0.1$. Numerical averages are taken over $10^6$ trajectories of a particle initially at the origin and oriented in the $x$-direction. The qualitative similarity between this figure and \f{figs1} emphasizes that isolated on-lattice and off-lattice active Brownian particles move in a similar fashion.}
    \label{figs3}
 \end{figure*} 
 
  The mean-squared displacement for $0 < \lambda <1$ reads
 \bea
 \label{msd_active3}
 \av{\rr_N \cdot \rr_N } &=& \av{\sum_{i=1}^N \Delta {\bm r }_i \cdot \sum_{i=j}^N\Delta \r_j} \nonumber \\
&=&\sum_{i=1}^N \av{\Delta \r_i \cdot \Delta \r_i} + 2 \sum_{i=1}^{N-1} \sum_{j=i+1}^N \av{\Delta \r_i \cdot \Delta \r_j} \nonumber \\
&=&N \frac{v_++2 v_0 + v_-}{\Sigma} \nonumber \\&+&2 \ell_0^2 \frac{1}{(1-\lambda)^2} (N(1-\lambda)-1 + \lambda^N  ),
 \eea
implying a long-time effective diffusivity
 \bea
 D_{\rm eff} &\equiv& \lim_{N \to \infty} \frac{1}{4N} \av{\rr_N \cdot \rr_N } \\  \label{deff2} &=&  \frac{1}{4} \frac{v_++2 v_0 + v_-}{\Sigma}+\frac{\ell_0^2}{2} \frac{1}{1-\lambda}.
 \eea
Thus an isolated active on-lattice Brownian particle behaves in a similar way to its off-lattice counterpart: it moves ballistically at short times, possesses a mean net displacement that is non-vanishing, and is effectively diffusive at long times.

 Collecting results for all three cases we have 
 \begin{widetext}
 \bea
   \av{\rr_N \cdot \rr_N }=  \left\{ 
  \begin{array}{cc} % brackets may be (...), [...], \{...\}, or left out
      N(N-1) \ell_0^2+N (v_++2 v_0+v_-)/\Sigma  &(\lambda = 1)\\
   {\rm  \eqq{msd_active3}} &(0< \lambda <1) \\
       N (v_++2 v_0+v_-)/\Sigma  &(\lambda = 0),
   \end{array}
   \right.
   \eea
    \end{widetext}
   for on-lattice driven, active, and Brownian matter, respectively.
   
 In \f{figs3} we confirm these analytic results numerically: an on-lattice active Brownian particle moves ballistically at short times, possesses a mean net displacement that is non-vanishing, and is effectively diffusive at long times, just like its off-lattice counterpart.
  
 \section{Flux-balance argument to estimate onset of phase separation}
 \label{fb}
 
To estimate when phase separation should occur on lattice we construct a kinetic-theory argument, following the argument used in Ref.\c{redner2013structure} to predict phase separation in an off-lattice model of active matter. Consider a simulation box containing a dense phase of density (number of particles per unit area) $\phi_{\rm d}$, and a gas of density $\phi_{\rm g}$. Consider a planar interface between these two phases, and focus on the net rate of departure and arrival of particles, at the interface, due to the persistent component of particle motion. We ignore the diffusive component of particle motion, because diffusion alone does not cause clusters to form.
 
 {\em Departure -- } Particles arriving at the interface will initially point into the dense phase. In order to leave, they must rotate so that they point in the opposite direction. The characteristic number of steps required for a particle to point in the direction opposite its arrival is $k = 2 \sum_{m=1}^\infty m\, 2^{-m} = 4$. The number of particles leaving the interface in this interval is $N_{\rm off} \sim l_{\rm int} \times \phi_{\rm d}$ (for $v+ \gg D_{\rm rot}$), where $l_{\rm int}$ is the length of the interface.
 
 {\em Addition -- } The characteristic number of steps made by an isolated particle, as a trapped particle makes a single rotational step, is $\Sigma/(2 D_{\rm rot}) \equiv 1/(1-\lambda)$. Thus an isolated particle makes $S = k/(1-\lambda)$ steps in the characteristic time taken for a trapped particle to become free. From the results of the previous section, only those isolated particles within a distance $|\av{\rr_{S}}| = \ell_0 (1-\lambda^{S})/(1-\lambda)$ can reach the interface in $S$ steps (recall that $\ell_0 \equiv (v_+-v_-)/\Sigma$). We then estimate that $N_{\rm on}$ is
 \beq
 N_{\rm on} \sim \frac{\phi_{\rm g}}{4} \times l_{\rm int} \times \ell_0 \frac{1-\lambda^{S}}{1-\lambda}.
 \eeq 
 The factor of $1/4$ comes from the fact that only $1/4$ of particles will, on average, point toward the interface.
 
 {\em Flux balance -- } Equating $N_{\rm off}$ and $N_{\rm on}$ we get
 \beq
 \label{ratio}
 \frac{\phi_{\rm d}}{\phi_{\rm g}} = \frac{v_+-v_-}{8 D_{\rm rot}} (1-\lambda^{k/(1-\lambda)}).
 \eeq
This relation contains the drift velocity of the particle and its rotational diffusion constant. As in the previous section it is natural to define the P\'{e}clet number 
  \beq
 \label{pec}
 {\rm Pe} \equiv \frac{v_+-v_-}{2D_{\rm rot}}.
 \eeq
 We also define $\kappa \equiv (1-\lambda^{k/(1-\lambda)})/4$, in which case \eq{ratio} reads
  \beq
 \label{ratio2}
 \frac{\phi_{\rm d}}{\phi_{\rm g}} = {\rm Pe} \, \kappa.
 \eeq
We shall assume that the dense phase has density $\phi_{\rm d} \approx 1$. 
 
{\em Mass conservation -- } There are $N_{\rm tot} = \phi A$ particles in the simulation box, $A$ being the simulation box area. Let there be $N_{\rm d} = \phi_{\rm d} A_{\rm d} \approx A_{\rm d}$ particles in the dense phase and $N_{\rm g} = \phi_{\rm g} (A - A_{\rm d})$ particles in the gas phase, where $A_{\rm d}$ is the area of the simulation box taken up by the dense phase. Mass conservation implies
 \beq
 \label{lever}
A_{\rm d} + \phi_{\rm g} (A - A_{\rm d}) = \phi A.
 \eeq
 We use \eqq{ratio2} to eliminate $\phi_{\rm g} = ({\rm Pe} \, \kappa)^{-1}$ from \eqq{lever}. We eliminate $A$ and $A_{\rm d}$ from the same equation in favor of $f \equiv A_{\rm d}/(\phi A)$, the fraction of particles in the solid phase. We get
 \beq
 \label{eff}
 f=\frac{{\rm Pe} \, \kappa-1/\phi}{{\rm Pe} \, \kappa-1}.
 \eeq
From \eqq{eff} we see that phase separation is only possible if ${\rm Pe} >1/(\kappa \phi)$. Thus infinite Pe is required to induce phase separation in the limit of vanishing packing fraction. For large ${\rm Pe}$ we have $f \to 1$, i.e. all particles will be in the dense phase. (In the limit of large $v_+$ we have $\kappa  \to (1-{\rm e}^{-4})/4 \approx 0.245$.)

The contour $f=1/2$ from \eqq{eff} is plotted against simulation data in Fig. 2. The correspondence shown there indicates that a flux-balance argument can describe the essence of motility-induced phase separation for on-lattice active matter, just as it does off lattice.

 \clearpage

\onecolumngrid

\setlength{\parskip}{0.25cm}%
\setlength{\parindent}{0pt}%

\section{Supplemental figures}
\begin{figure*}[b]
\includemovie[poster= 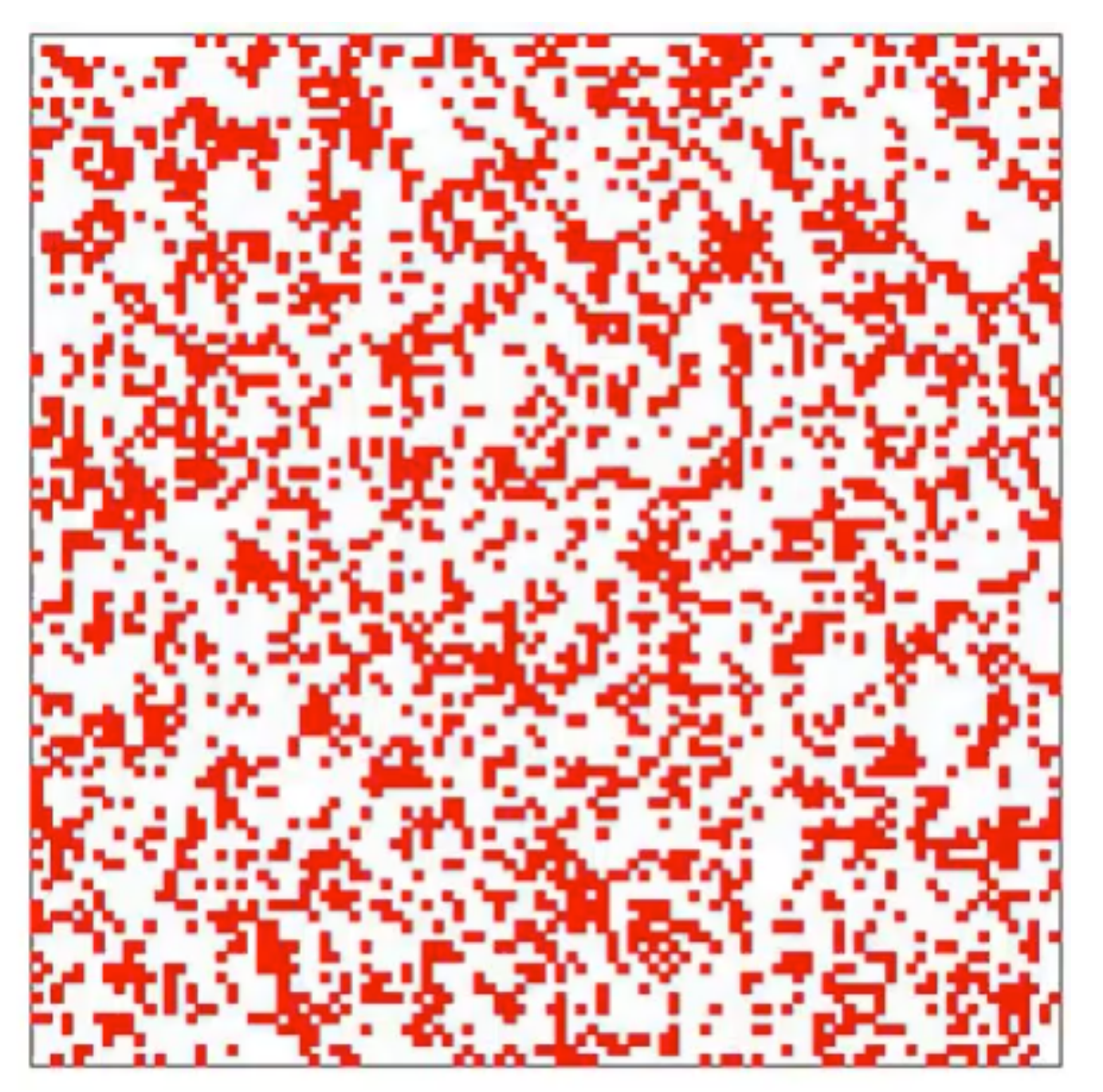, mouse]{0.5\linewidth}{0.5\linewidth}{mov1.mp4}
\caption{\label{fig_movie} Movie of phase separation and cluster dynamics: click to animate (when viewing document in Adobe Reader).}
\end{figure*}

\clearpage
   \begin{figure*}[] 
    \centering
    \includegraphics[width=\linewidth]{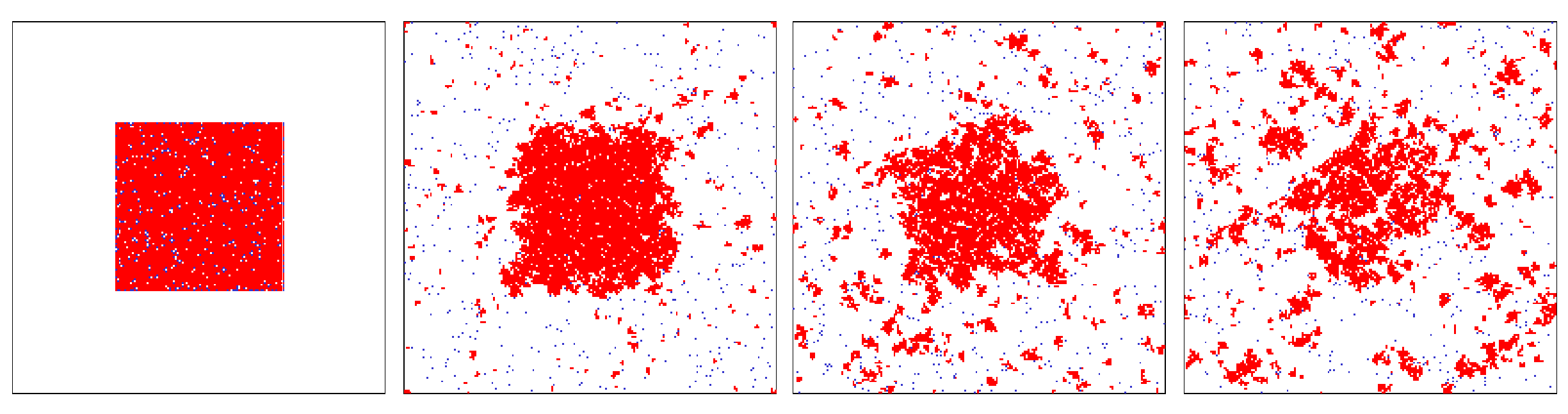} 
    \caption{When lateral motion is not possible (here $v_-=v_0=0$), phase separation does not occur. A pre-built compact cluster of particles dissolves into several smaller ones. ($v_+=25$, $\phi=1/5$). When lateral motion is restored, phase separation is made possible; see Fig. 1 of the main text.}
    \label{fig3}
 \end{figure*}

 \end{document}